\documentclass[aps,prl,showpacs,footinbib,superscriptaddress,twocolumn,longbibliography]{revtex4-2}
\usepackage{amsfonts}
\usepackage{amsmath}
\usepackage{multirow}
\usepackage{txfonts}
\usepackage{amssymb}
\usepackage{amsbsy}
\usepackage{graphicx}
\usepackage{epstopdf}
\usepackage{color}
\usepackage{braket} 
\usepackage{mathdots} 
\usepackage{booktabs}
\usepackage{indentfirst}
\usepackage{hyperref}
\usepackage{diagbox}
\usepackage{lipsum}
\usepackage{booktabs}
\usepackage{bibentry}
\hypersetup{ colorlinks=true, linkcolor=blue, anchorcolor=blue, citecolor=blue,urlcolor=blue}
\usepackage{filecontents}

\begin{document}
\title{Asymptotic Exceptional Steady States in Dissipative Dynamics}

\author{Yu-Min Hu}
\affiliation{Max Planck Institute for the Physics of Complex Systems, N\"{o}thnitzer Str. 38, 01187 Dresden, Germany}

\author{Jan Carl Budich}
\altaffiliation{jan.budich@tu-dresden.de}

\affiliation{Max Planck Institute for the Physics of Complex Systems, N\"{o}thnitzer Str. 38, 01187 Dresden, Germany}
\affiliation{Institute of Theoretical Physics, Technische Universit\"at Dresden and W\"urzburg-Dresden Cluster of Excellence ct.qmat, 01062 Dresden, Germany}

\begin{abstract}
Spectral degeneracies in Liouvillian generators of dissipative dynamics generically occur as exceptional points, where the corresponding non-Hermitian operator becomes nondiagonalizable. Steady states, i.e., zero modes of Liouvillians, are considered a fundamental exception to this rule since a no-go theorem excludes nondiagonalizable degeneracies there. Here, we demonstrate that the crucial issue of diverging timescales in dissipative state preparation is largely tantamount to an asymptotic approach toward the forbidden scenario of an exceptional steady state in the thermodynamic limit. With case studies ranging from NP-complete satisfiability problems encoded in a quantum master equation to the dissipative preparation of a symmetry protected topological phase, we reveal the close relation between the computational complexity of the problem at hand, and the finite size scaling toward the exceptional steady state, exemplifying both exponential and polynomial scaling. Formally treating the weight $W$ of quantum jumps in the Lindblad master equation as a parameter, we show that exceptional steady states at the physical value $W=1$ may be understood as a critical point hallmarking the onset of dynamical instability.           

\end{abstract}

\maketitle
Quantum state preparation remains a formidable challenge due to diverging timescales for large systems, with implications reaching far beyond its natural habitat of quantum science. In particular, the solution to hard problems of general interest such as NP-complete satisfiability (SAT) problems \cite{cook2023complexity,karp2009reducibility,Mezard2002random} and their quantum counterparts known as QMA-complete problems \cite{kitaev2002classical,kempe2006complexity,aharonov2008adiabatic} may be readily viewed as quantum state preparation tasks, e.g. solved by preparing the ground state of a suitable local parent Hamiltonian in a quantum simulator \cite{buluta2009quantum,cirac2012goals,Georgescu2014Quantumsimulation,Daley2022Practical}.For the common approach of adiabatic quantum computing \cite{farhi2001quantum,Das2008Colloquium,Albash2018Adiabatic}, the central obstruction is  a (discontinuous) quantum phase transition along the path to nontrivial ground states, manifesting as a (exponentially) divergent preparation time, known as critical slowdown
\cite{Young2010firstorder,Jorg2010Firstorder,altshuler2010anderson,Amin2009firstorder,Dickson2011does,farhi2009quantum,Werner2023Bounding}. 

Dissipative state preparation aims at addressing this challenge by considering incoherent processes as a resource \cite{Kraus2008Preparation,diehl2008quantum,verstraete2009quantum,diehl2011topology,bardyn2013topology, Budich2015dissipativechern, Moshe2019dissipation_topology, Liu2021dissipative, Yang2023dissipative, Bandyopadhyay2020Dissipative, Zhou2021SPT, lin2013dissipative, Reiter2016Scalable, Kastoryano2011Dissipative}. In this way, a targeted many-body state is approached independently of initial conditions as a steady state of dissipative dynamics. There, a central equation of motion is the Lindblad master equation
\cite{gorini1976completely,lindblad1976generators}
\begin{equation}\label{eq:lindblad}
    \frac{\mathrm{d}}{\mathrm{d}t}\rho=\mathcal{L}_W[\rho]=-i(H_{\text{nH}}\rho - \rho H_{\text{nH}}^\dagger)+  W\sum_{\mu}L_\mu\rho L_\mu^\dagger,
\end{equation}
governing the dynamics of the density matrix $\rho$ of a system weakly coupled (Born) to a bath with negligible memory (Markov). Incoherent processes reflecting the influence of the bath are described by quantum jump operators $L_\mu$, while $H_{\text{nH}}=H-(i/2)\sum_{\mu} L_\mu^\dagger L_\mu$ is the effective non-Hermitian (NH) Hamiltonian combining the Hermitian system Hamiltonian $H$ with the anti-Hermitian damping enacted by the bath. We emphasize that the weight $W$ of the quantum jump term is physically fixed to $W=1$, but for our subsequent analysis, it will be fruitful to formally consider $W$ as a parameter \cite{Dalibard1992Wave-function,Plenio1998RMP,daley2014quantum,Garrahan2010Thermodynamics,Carollo2019Unraveling,supp_mater}.

\begin{figure}[t]
    \centering
    \includegraphics[width=8.5cm]{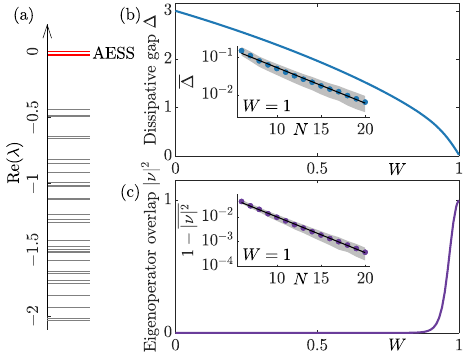}
    \caption{(a) Liouvillian spectrum ($W=1$ in Eq. \eqref{eq:lindblad}) for a satisfiable 3SAT instance with a unique solution, $N=14$ variables, and $M=\text{round}(\alpha_cN)$ clauses, where the satisfiability threshold $\alpha_c\approx4.267$. The two red states form the asymptotic exceptional steady state (AESS). (b,c) Liouvillian gap $\Delta$ and eigenoperator overlap $|\nu|^2$ as a function of $W$ (other parameters as in (a)).  The insets in (b,c) show the mean values of  $\overline\Delta$  and $\overline{|\nu|^2}$ at $W=1$ change with the number of variables $N$, averaging over $10^3$ random instances for each $N$ with $M=\text{round}(\alpha_cN)$. All instances have a unique solution. Shadow regions indicate the standard deviation among instances.}
    \label{fig1}
\end{figure}

Importantly, the aforementioned critical slowdown persists in the dissipative context, namely as an asymptotic degeneracy at eigenvalue zero of  the Liouvillian $\mathcal{L}_{W=1}$ [cf. Fig. \ref{fig1}(a)], raising the question of the nature of this degeneracy. On one hand,  it is known that NH degeneracies generically occur in the form of nondiagonalizable (or exceptional) points  (EP)\cite{heiss2004exceptional,berry2004physics,heiss2012physics, el2018non,ozdemir2019parity,miri2019exceptional}, such as Liouvillian exceptional points \cite{Minganti2019Quantum, Arkhipov2020Liouvillian, Minganti2020Hybrid, Khandelwal2021signatures, Chen2021Quantum,Zhang2022Dynamical, Chen2022Decoherence, zhou2023accelerating, Bu2023Enhancement}, while diagonalizable degeneracies would require additional fine-tuning \cite{state_EP}.  On the other hand, a rigorous no-go theorem precludes degenerate steady states (i.e., Liouvillian zero modes) from forming an EP \cite{Minganti2018spectral}. 

\begin{figure}[t]
    \centering
    \includegraphics[width=8.5cm]{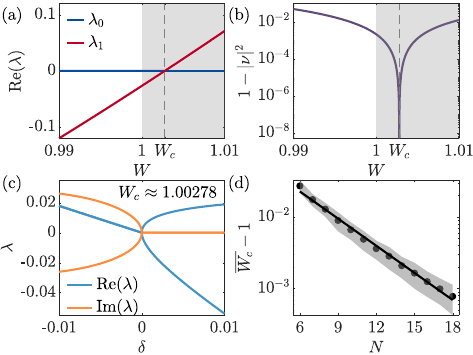} 
     \caption{Eigenvalues (a) of steady and metastable states and their eigenoperator overlap (b) over $W$ [same instance as in Fig. \ref{fig1}]. The shaded region represents $W>1$, where $\text{Re}(\lambda_1)>0$ renders the system dynamically unstable in the thermodynamic limit. We observe an EP at $W_c\approx 1.00278$. (c) Perturbed eigenvalues for $\mathcal{L}_{W_c}+\mathcal{L}_{W_c}^{\text{pert}}$, where $\delta$ is the perturbation strength. (d) The mean $W_c$ decays with $N$. Each point is averaged over $10^3$ instances with one solution and the shaded region marks the standard deviation between instances.}
    \label{fig2}
\end{figure}

\begin{figure}[t]
    \centering
    \includegraphics[width=8.5cm]{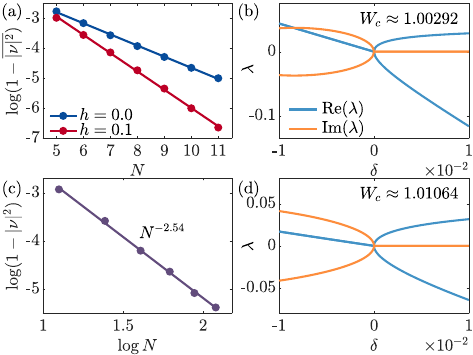}
    \caption{AESS in quantum systems. (a) The case of including a modified transverse field [Eq. \eqref{eq:PXP}] in dissipative 3SAT solver. We set $M=\text{round}(\alpha_c N)$ and $W=1$. All points are obtained by averaging over $10^3$ satisfiable instances with a unique solution. (b)  The perturbed eigenvalues of $\mathcal{L}_{W_c}+\mathcal{L}_{W_c}^{\text{pert}}$ for a uniquely solvable 3SAT instance with $N=11$. (c) Eigenoperator overlap of $\mathcal{L}_1$ for preparing the AKLT state [Eq. \eqref{eq:AKLT}]. (d) The perturbed eigenvalues of $\mathcal{L}_{W_c}+\mathcal{L}_{W_c}^{\text{pert}}$ for AKLT state preparation with $N=7$. }
    \label{fig3}
\end{figure}

Here, we report on the discovery of {\em{asymptotic exceptional steady states}} (AESS) as the generic form of an asymptotic degeneracy associated with critical slowdown in dissipative dynamics. In essence, the forbidden scenario of a steady state EP is approached in the thermodynamic limit (see Fig. \ref{fig1} for a paradigmatic example), while the exact EP is pushed (infinitesimally) away from the physical value $W=1$ [cf. Eq. \eqref{eq:lindblad}] to $W_c > 1$  in any finite system (see Fig. \ref{fig2}). Furthermore, we reveal that the complexity of a state preparation task at hand naturally determines the finite size scaling properties of AESS, as we exemplify by two concrete systems of distinct scientific interest. First, for the 3SAT problem (cf. Fig. \ref{fig1}), the AESS is approached exponentially in system size $N$, reflecting the NP-completeness of the problem. Second, for a Lindbladian with a symmetry-protected topological phase as a steady state, we find that an AESS is approached with polynomial scaling, reflecting the continuous topological quantum phase transition separating the steady state from a trivial initial state in symmetry-preserving dynamics [cf. Fig. \ref{fig3}(c) below].

\emph{AESS in dissipative state preparation.--} The central idea of the dissipative state preparation paradigm is to target a pure state $\rho_t=\ket{\psi_t}\bra{\psi_t}$ by encoding it into the steady-state subspace of a physical open quantum system, i.e., $\mathcal{L}_1[\rho_t]=0$. In our present approach, we \textcolor{black}{first} set $H=0$ for simplicity and consider jump operators of the form $L_\mu=O_\mu P_\mu$. \textcolor{black}{The case with $H\ne0$ will be discussed further below.} Here, the projector $P_\mu$ satisfies $P_\mu\ket{\psi_t}=0$ and the operator $O_\mu$ maps unwanted states into the subspace containing $\ket{\psi_t}$. In other words, $\ket{\psi_t}$ lies in the decoherence-free subspace \cite{Lidar1998decoherencefree} while jump operators dissipate undesirable states. 

These jump operators lead to an anti-Hermitian effective Hamiltonian 
\begin{equation}\label{eq:HnH}
H_{\text{nH}}=-(i/2)\sum_\mu P_\mu O_\mu^\dagger O_\mu P_\mu=-i H^\prime,
\end{equation}
such that $\ket{\psi_t}$ is a frustration-free ground state of the Hermitian $H^\prime$.  Neglecting the quantum jump terms (i.e., $W=0$), $\mathcal{L}_{0}[\rho]=-\{H^\prime,\rho\}$ generates the deterministic imaginary time evolution of $H^\prime$ \cite{mcardle2019variational,nishi2021implementation}. While the finite spectrum gap at $W=0$ [see Fig. \ref{fig1}(b)] suggests convergence to the solution in a time scaling at most linearly in system size \cite{Mori2020resolving,Haga2021Liouvillian}, we emphasize that a direct experimental implementation of $W=0$ requires postselection of processes (quantum trajectories) without quantum jump events, the probability of which is attenuated exponentially over time.

As for the physical Liouvillian $\mathcal{L}_{W=1}$, a key quantity to characterize critical slowdown is the relaxation time $\tau$, which is determined by the Liouvillian spectrum.  A (diagonalizable) Liouvillian $\mathcal{L}_W$ satisfies $\mathcal{L}_W[r_i]=\lambda_{i}r_i$ and $\mathcal{L}_W^\dagger[l_i]=\lambda_{i}^*l_i$, respectively \footnote{Our work focuses on the spectrum near the steady state which can be generically assumed to be diagonalizable. A nondiagonalizable Liouvillian can be treated similarly by replacing the eigenvalues and eigenstates of decaying modes with those of the corresponding Jordan blocks.}. Here, $\lambda_{i}$ is the $i$th eigenvalue, ordered by their real parts $\text{Re}(\lambda_{0})\ge\text{Re}(\lambda_{1})\ge\text{Re}(\lambda_{2})\ge\cdots\ge \text{Re}(\lambda_{D^2-1})$, where $D$ is the Hilbert space dimension, and ${r_i}$ (${l_i}$) are the right (left) eigenoperators whose Hilbert-Schmidt inner product satisfies the biorthonormal relation  $\text{Tr}[l^\dagger_i r_j]=\delta_{ij}$. At the physical point ($W=1$), the eigenmodes with $\lambda_{i}=0$ correspond to the steady-state subspace. To clarify the concept of AESS, we assume a unique steady state of $\mathcal{L}_1$ throughout the Letter, which is the most common case in dissipative state preparation. Then, $\lambda_0=0$ and $\text{Re}(\lambda_1)<0$ at $W=1$. Consequently, it is the Liouvillian gap $\Delta=|\text{Re}(\lambda_0-\lambda_1)|$ between the steady state $r_0$ and the slowest relaxation mode $r_1$ that reveals the relaxation time $\tau\sim\Delta^{-1}$. For generalizations in which the AESS may coexist with multiple steady states, see the Supplemental Material (SM) \cite{supp_mater}.

For nontrivial dissipative preparation tasks, including the preparation of topological states and solution states to computationally hard problems, a divergent timescale in the thermodynamic limit is inherent, thus demanding asymptotic gap closing with increasing system size. This generically occurring critical slowdown is simply the manifestation of many-body complexity in dissipative state preparation. Nevertheless, $r_0$ and $r_1$ in a gapless Liouvillian are never strictly degenerate because of the finite-size gap that is typically present in all physically accessible systems. While a no-go theorem precludes EPs for exactly degenerate steady states at $W = 1$  \cite{Minganti2018spectral}, the asymptotic (avoided) degeneracy in gapless Liouvillians does not fall into this category. As EPs are the most generic form of NH spectral degeneracy \cite{state_EP}, we argue that without fine tuning or additional symmetry, any asymptotic steady-state degeneracy must also lie close to an exact EP. This EP can be revealed by considering an extended (and thus mathematically generic) parameter space (as is explicated below by considering $W \ne 1$, see Fig. \ref{fig2}), where the no-go theorem no longer applies. Thus, the AESS where both eigenvalues and eigenstates between the steady state and slowest relaxation mode asymptotically coalesce with increasing system size, indeed represents the typical manifestation of critical slowdown in dissipative state preparation.

To quantitatively describe the properties of AESS, we define the eigenoperator overlap between $r_0$ and $r_1$:
\begin{equation}\label{eq:eta}
    \nu=\text{Tr}[\hat r_0^\dagger\hat r_1],
\end{equation}
where we define the normalized eigenoperators ${\hat r_i}={r_i}/||r_i||$ with the Hilbert-Schmidt norm $||\rho||=\sqrt{\text{Tr}[\rho^\dagger\rho]}$.  $|\nu|^2=0$  ($|\nu|^2=1$) reveals that these two states are orthogonal (parallel) to each other. In this sense, an AESS is characterized by $\Delta \to 0$ and $|\nu|^{2} \to 1$ in a physical system (at $W = 1$) being simultaneously approached in the thermodynamic limit.

Whenever $|\nu|^2 \ne1$, an orthonormal operator basis between $r_0$ and $r_1$ can be constructed:  ${\tilde r_0}=\hat r_0$ and ${\tilde r_1}=({\hat r_1}-\nu\hat r_0)/({1-|\nu|^2})^{1/2}$, in which the Liouvillian superoperator $\mathcal{L}_W$ becomes \cite{supp_mater}   
\begin{equation}\label{eq:leff}
   \mathcal{L}_{W}^{\text{eff}}=\begin{pmatrix}
      0 & \lambda_1 \frac{\nu}{\sqrt{1-|\nu|^2}} \\ 0 & \lambda_1
    \end{pmatrix}.
\end{equation}
Here, we always have $\lambda_0=0$ for generic $W$ since $r_0=\ket{\psi_t}\bra{\psi_t}$ by assumption lies in the decoherence-free subspace of the preparation protocol. As an effective description of the AESS, $\lambda_1\to 0$ and $|\nu|^2\to1$ are simultaneously approached by either taking the thermodynamic limit [Fig. \ref{fig1}(b,c)] or slightly tuning $W$ away from the physical value $W=1$ [Fig. \ref{fig2}(a,b)]. In this sense, the effective matrix $\mathcal{L}_W^{\text{eff}}$ asymptotically approaches the generic Jordan-block structure of an EP \footnote{The effective description $\mathcal{L}_W^{\text{eff}}$ in Eq.\eqref{eq:leff} indicates that an open quantum system at $W=1$ cannot approach the AESS when only one of the limits $\lambda\to0$ and $|\nu|^2 \to 1$ is achieved. Examples of those cases are presented in the SM\cite{supp_mater}.}.

Having introduced the general notion of AESS in dissipative dynamics, we now demonstrate the characteristic dependence of its scaling behavior on the class of state preparation problem with paradigmatic examples.

\emph{Dissipative 3SAT solver.--} We first consider the NP-complete 3SAT problem.  With $N$ Boolean variables $x_1,x_2,\cdots,x_N$, we define $M$ disjunction clauses, each containing three variables or their negations (e.g., $C_m=x_{m_1}\lor \neg {x}_{m_2}\lor x_{m_3}$). The 3SAT problem asks whether a conjunction of clauses $C_1\land C_2\land\cdots\land C_M$ can be satisfied by assigning TRUE (1) or FALSE (0) to each binary variable. By mapping TRUE and FALSE of a variable $x_n$ to the states $\ket{1}$ and $\ket{0}$ of a qubit $\sigma_n$, the 3SAT problem is converted into finding the ground state of a 3-local Ising-like Hamiltonian $H_{\text{3SAT}}=\sum_{m=1}^MP_m$ \cite{lucas2014ising}. The projector $P_m$ acts on three qubits involved in clause $C_m$ and assigns a unit energy penalty to unsatisfied clauses. For example, $C_m=x_{m_1}\lor \neg {x}_{m_2}\lor x_{m_3}$ corresponds to $P_m=(1-\sigma_{m_1}^z)(1+\sigma_{m_2}^z)(1-\sigma_{m_3}^z)/8$, which punishes the configuration $\ket{010}$ of the three participating variables. Here, $\sigma_n^{x,y,z}$ are Pauli operators for the $n$th qubit, with $\sigma_n^z\ket{1}=\ket{1}$ and  $\sigma_n^z\ket{0}=-\ket{0}$. The solution to a satisfiable 3SAT instance is encoded in a zero-energy ground state $\ket{\psi_{\text{sol}}}$ of $H_{\text{3SAT}}$, where $P_m\ket{\psi_{\text{sol}}}=0$ and $\ket{\psi_{\text{sol}}}$ represents a bit string. This maps the problem to the standard state preparation task of preparing the ground states of a target Hamiltonian $H_{\text{3SAT}}$. Yet, the NP-complete nature of 3SAT indicates the generic hardness of completing this task.

To harness the dissipative state preparation framework, we design dissipative processes for each clause $C_m$ through three jump operators:
\begin{equation}\label{eq:classical_3SAT}
    L_{m,\alpha}=\sigma_{m_\alpha}^xP_m,\quad \alpha=1,2,3.
\end{equation}
These $L_{m,\alpha}$ dissipate the unsatisfiable configuration of clause $C_m$ and rotate qubit $m_\alpha$ into its satisfiable subspace. Therefore, the solution to a satisfiable 3SAT instance is given by the steady state $r_0=\ket{\psi_{\text{sol}}}\bra{\psi_{\text{sol}}}$ of  the corresponding $\mathcal{L}_1$. Remarkably, Eq. \eqref{eq:classical_3SAT} leads to $H^\prime=3H_{\text{3SAT}}/2$ [cf. Eq. \eqref{eq:HnH}], indicating that the imaginary time evolution of $H_{\text{3SAT}}$ is achieved by $ \mathcal{L}_{W=0}$. While the evolution of $\mathcal{L}_0$ approaches $\ket{\psi_{\text{sol}}}$ at a timescale that is at most linear in $N$ due to a finite (and constant with $N$) gap of $H_{\text{3SAT}}$ [Fig. \ref{fig1}(b)], the hardness of the 3SAT problem hides in postselecting an exponentially small subset of quantum trajectories to effectively realize $W=0$.  An interesting question relating to our present discussion of AESS is: How does the complexity of 3SAT problem manifest in the spectrum of $\mathcal{L}_W$ as a function of $W$, in particular at the natural physical value $W=1$?

To examine the spectrum of $\mathcal{L}_W$, we consider satisfiable 3SAT instances with a unique solution. Similar results with multiple solutions are presented in the SM \cite{supp_mater,unsat_case}. Hard satisfiable instances are generated using the method from Ref. \cite{Barthel2002Hiding}. The clause-to-variable ratio $M/N$ is set to the satisfiability threshold $\alpha_c\approx 4.267$, below (above) which a generic instance is satisfiable (unsatisfiable)  \cite{Mezard2002random}. In the computational basis with $H=0$, the structure of Eq. \eqref{eq:classical_3SAT} allows us to focus on the classical dynamics in the diagonal part of $\rho$,  thus reducing the effort in diagonalizing $\mathcal{L}_W$ \cite{supp_mater}.

The spectrum of $\mathcal{L}_1$ for a typical satisfiable instance is presented in Fig. \ref{fig1}(a). Besides the steady state $r_0=\ket{\psi_{\text{sol}}}\bra{\psi_{\text{sol}}}$, we observe a metastable state $r_1$ that is well separated from other damping modes, with eigenvalue close to zero. As shown in Fig. \ref{fig1}(b), this state has a small Liouvillian gap at $W=1$, in contrast to the finite gap at the imaginary time evolution point $W=0$. Importantly, the mean Liouvillian gap $\overline{\Delta}$ at $W=1$, averaged over $10^3$ satisfiable instances, exhibits an exponential decay as $N$ increases, indicating the exponential runtime of a dissipative 3SAT solver. 

The most interesting result appears in the eigenoperator overlap $\nu$  defined in Eq. \eqref{eq:eta}. As shown in Fig. \ref{fig1}(c), we find that $|\nu|^2=0$ at $W=0$, as the two eigenoperators correspond to eigenstates of a Hermitian Hamiltonian $H_{\text{3SAT}}$. However, $|\nu|^2$ grows to a value that is extremely close to 1 at $W=1$. The inset of Fig. \ref{fig1}(c) further demonstrates that the averaged quantity $1-\overline{|\nu|^2}$ at $W=1$  decays exponentially as $N$ increases. These numerical findings imply that the metastable and steady states asymptotically approach an EP in the thermodynamic limit ($N\to+\infty)$, leading to the appearance of AESS.

In a finite system, an exponentially small finite Liouvillian gap at $W=1$ prevents two nearly degenerate states  from becoming an exact EP. However, a slight increase of $W$ to a critical value $W_c>1$ makes $\Delta=0$ and $|\nu|^2=1$,  indicating that two states coalesce to an exact EP in an extended parameter space [Fig. \ref{fig2}(a-b)]. This shows that the effective Liouvillian in Eq. \eqref{eq:leff} approaches a Jordan block as $W\to W_c$.  Notably, the EP at $W=W_c$ signifies a transition into a dynamically unstable region where $\lambda_1>0$ and the steady state becomes ill-defined \footnote{When $\lambda_0$ and $\lambda_1$ in Fig. \ref{fig2}(a) cross at $W=W_c$,  we keep denoting the eigenvalue of the solution state $\ket{\psi_{\text{sol}}}\bra{\psi_{\text{sol}}}$ with $\lambda_0$ for simplicity. This keeps Eq. \eqref{eq:leff} valid at both $W>W_c$ and $W<W_c$ without exchanging basis vectors.}.  Moreover, the critical value $W_c$ converges to $1$ in the thermodynamic limit, whose mean deviation $\overline{W_c}-1$ is compatible with an exponential decay with system size [Fig. \ref{fig2}(d)].  These results demonstrate that the physical AESS at $W=1$ is extremely close to an exact EP. 

To further demonstrate that the AESS at $W=1$ is near an exact EP at $W=W_c$, we perturb the spectrum of $\mathcal{L}_W$ by adding a perturbative Liouvillian $ \mathcal{L}_{W}^{\text{pert}}[\rho]=\delta\sum_{n=1}^N(WL_n\rho L_n^\dagger-\frac{1}{2}\{L^\dagger_n L_n,\rho\})$ with $L_n=\sigma_n^-$, where  $\delta$ is the perturbation strength. Since a solution $\ket{\psi_{\text{sol}}}$ to a random satisfiable instance is unlikely to be a dark state for all $L_n=\sigma_n^-$, this perturbation erases the solution in the steady state. While a physical dissipative dynamics requires  $\delta>0$ and $W=1$, we mathematically consider both $\delta>0$ and $\delta<0$  at $W=W_c>1$. As shown in Fig. \ref{fig2}(c), the perturbed eigenvalues are purely real when $\delta>0$ and form complex-conjugated pairs when $\delta<0$, undergoing a parity-time symmetry breaking at $\delta=0$ \cite{Bender1998Real,bender2002generalized,mostafazadeh2002pseudo,mostafazadeh2001pseudo}. This is a clear signature  of  an exact EP at $\delta=0$ and $W=W_c$.

\emph{Quantum AESS.--} While the above dissipative 3SAT solver is governed by classical dynamics, our arguments for the occurrence of AESS also apply to genuine quantum dynamics. To exemplify this case, we consider two quantum models: a modified dissipative 3SAT solver with $H\ne 0$, and a dissipative 
preparation of the Affleck–Kennedy–Lieb–Tasaki (AKLT) state \cite{Affleck1987AKLT_model}. 

As a modified 3SAT solver, we add a quantum Hamiltonian $H_X$ to Eq. \eqref{eq:classical_3SAT}:
\begin{equation}\label{eq:PXP}
    H_X=h\sum_{m\ne m^\prime}^M\sum_{n\in \text{Ind}_{m,m^\prime}}P_{m,m^\prime}\sigma_n^xP_{m,m^\prime}
\end{equation}
where $P_{m,m^\prime}=I-(I-P_m)(I-P_{m^\prime})$, $h$ is the interaction strength, and $\text{Ind}_{m,m^\prime} $ contains the qubit indices involved in clauses $C_m$ and $C_{m^\prime}$. $H_X$ describes a transverse field that only generates dynamics in the subspace orthogonal to $\ket{\psi_{\text{sol}}}$.  Therefore, $\ket{\psi_{\text{sol}}}\bra{\psi_{\text{sol}}}$ is still the steady state at $W=1$. Figure \ref{fig3}(a) indicates that the AESS still exists  in quantum dynamics, where both the eigenvalues and eigenoperators of steady and metastable states for $\mathcal{L}_1$ coalesce in the thermodynamic limit. Additionally, an exact EP appears at $W=W_c>1$, which is confirmed by the perturbed eigenvalues [Fig.\ref{fig3}(b)] at $W=W_c$ after adding a perturbation $ \mathcal{L}_{W_c}^{\text{pert}}[\rho]=\delta\sum_{n=1}^N(W_cL_n\rho L_n^\dagger-\frac{1}{2}\{L^\dagger_n L_n,\rho\})$ with $L_n=\sigma_n^x$.  

Both 3SAT solvers eventually stabilize a classical steady state. We now demonstrate that an AESS can also occur as a nontrivial quantum state, such as the symmetry-protected topological AKLT state \cite{Kraus2008Preparation,Zhou2021SPT,Wang2023Dissipative}. We consider a one-dimensional $S=1$ spin system ($H=0$) with four types of jump operators:
\begin{equation}\label{eq:AKLT}
    L_{n,\alpha}=S^{\alpha}_{n}P_{n,n+1}^{(2)},  \quad L_{n,\alpha}^\prime=S^{\alpha}_{n+1}P_{n,n+1}^{(2)},
\end{equation}
where $\alpha=x,y$. $S_n^{x,y,z}$ are spin-1 operators.  $P_{n,n+1}^{(2)}$ projects two neighboring $S=1$ spins into the $S=2$ subspace \cite{Affleck1987AKLT_model}. We assume that there are $N$ spins and take periodic boundary conditions $S_1=S_{N+1}$. By construction, the AKLT state $\ket{\psi_{\text{AKLT}}}$ lives in the decoherence-free subspace since $P_{n,n+1}^{(2)}\ket{\psi_{\text{AKLT}}}=0$ for $n=1,\cdots, N$, thus giving the steady state $\ket{\psi_{\text{AKLT}}}\bra{\psi_{\text{AKLT}}}$  of $\mathcal{L}_1$. We find that the AESS still exists at $W=1$ [Fig. \ref{fig3}(c)] and is close to an exact EP at a slightly larger $W=W_c>1$. We also perturb the exact EP  by adding a perturbative Liouvillian  $ \mathcal{L}_{W_c}^{\text{pert}}[\rho]=\delta\sum_{n=1}^N(W_cL_n\rho L_n^\dagger-\frac{1}{2}\{L^\dagger_n L_n,\rho\})$ with $L_n=S_n^z$. The perturbed eigenvalues [Fig.\ref{fig3}(d)]  elucidate the spectral EP structure.

Interestingly, the eigenoperator overlap $1-|\nu|^2$ in Fig. \ref{fig3}(c) decays as a power law with increasing system size, distinct from the exponential decay observed in Figs.\ref{fig1} and \ref{fig3}(a) for 3SAT. These numerical findings further corroborate the close relation between the finite-size scaling of the AESS and the complexity of preparing target states. Preparing a symmetry-protected topological state in a symmetry-preserving fashion requires a polynomial time reflecting the obstruction by a continuous topological quantum phase transition, while an exponentially long time is expected to be required to solve the NP-complete 3SAT problem. 

\emph{Dynamical consequences of AESS.--} We now return to a more general discussion of the physical Liouvillian at $W=1$. Although a finite-size gap prevents a direct perturbative analysis, the AESS still plays a crucial role in the complexity of approaching the physical steady state. With the biorthonormal relation $\text{Tr}[l_i^\dagger r_j]=\delta_{ij}$ for $\mathcal{L}_1$,  an initial density matrix $\rho_{\text{ini}}$ follows the time evolution $\rho(t)=r_0 + \sum_{i\ge 1}e^{\lambda_it}c_ir_i$, where $c_i=\text{Tr}[l_i^\dagger \rho_{\text{ini}}]$. In the longtime regime, the evolution effectively becomes $\rho(t)=r_0 + e^{\lambda_1t}c_1r_1+o(e^{\lambda_2t})$. 

The eigenoperators $r_1$ and $l_1$ of AESS have the form: $r_1=r_0+\delta_r$ and  $l_1=\kappa(I-r_0)+\delta_l$, where $||\delta_r||/||r_1||$ and $||\delta_l||/||l_1||$ decay to $0$ and $\kappa\equiv \operatorname{Tr}[l_1]/(2^N-1)$ approaches $-1$ as $N$ increases \cite{supp_mater}.  Since $l_1$ is close to the projector $I-r_0$, a random initial state without prior knowledge of  $r_0=\ket{\psi_{\text{sol}}}\bra{\psi_{\text{sol}}}$ will typically lead to $c_1=\text{Tr}[l_1^\dagger \rho_{\text{ini}}]=-1+\epsilon$ with an exponentially small correction $\epsilon$. The longtime dynamics becomes $\rho(t) \approx (1 - e^{\lambda_1 t} + \epsilon e^{\lambda_1 t}) r_0 + e^{\lambda_1 t} (-1 + \epsilon) \delta_r$, intrinsically requiring a time scale $\tau \sim |\operatorname{Re}(\lambda_1)|^{-1}$ to distinguish the target state $r_0$. One possible way to accelerate the relaxation dynamics is to carefully choose an initial state such that $c_1=0$, which would lead to a much shorter relaxation time $\tau^\prime\sim |\text{Re}(\lambda_2)|^{-1}$ \cite{Carollo2021Exponentially}.  However, such shortcuts are not readily applicable to the AESS considered here: as numerically exemplified in the SM \cite{supp_mater} for the dissipative 3SAT solver, we find that $\kappa^{-1}l_1$ is positive semidefinite, with a single zero mode $\ket{\psi_{\text{sol}}}$ separated from other positive eigenvalues by a gap that is constant in $N$. This situation excludes the possibility of finding easily accessible initial states $\rho_{\text{ini}}$ such that $c_1=0$, as the only solution would be to (tautologically) start from the target state $r_0$.  This further corroborates that the dynamical obstruction to preparing nontrivial target states manifests in the scaling properties of AESS, thus reflecting the intrinsic complexity of dissipative state preparation.

\emph{Conclusion.--} We have revealed the existence of asymptotic exceptional steady states and discussed their intriguing connection to the computational complexity of state preparation tasks. The AESS is found to be near an exact EP in an extended parameter space and to play a vital role in relaxation dynamics to target states. \textcolor{black}{Moreover, we note that AESS exhibit a subtle relation to the steady-state phase diagram of a dissipative system. In particular, it is important to distinguish between critical steady states in the sense of quantum phases of matter and the criticality of Liouvillian spectra. The latter, which has the dynamical meaning of critical slowdown and has been the focus of our present discussion of AESS, contains information both about the target state and the way to approach it.} Another interesting future direction is to identify Hamiltonian or dissipative perturbations that are compatible with the targeted steady state but mitigate the scaling behavior of the AESS so as to speed up the state preparation task. Such perturbations may be seen as a dissipative counterpart to  quantum catalysts in adiabatic quantum computing \cite{Albash2018Adiabatic}.

\emph{Acknowledgment.--} We thank Emil J. Bergholtz, Sebastian Diehl, Tim Pokart, Joachim Schwardt, and Grigorii Starkov for helpful discussions. JCB acknowledges financial support  from the German Research Foundation (DFG) through the Collaborative Research Centre SFB 1143 (Project-ID 247310070), and the Cluster of Excellence ct.qmat (Project-ID 390858490).

\emph{Data availability.--}The data associated with this work are available in a Zenodo repository \footnote{Data repository available at:  \href{https://doi.org/10.5281/zenodo.17661045}{https://doi.org/10.5281/zenodo.17661045}}

\emph{Note added.--} While preparing this manuscript for submission, we became aware of a somewhat related preprint \cite{gu2025exploring} which discusses Liouvillian EPs in a few-level system with $W\ne1$, but without relating to the large $N$ phenomenon of AESS introduced in our present Letter.  

\bibliography{AEP_ref}

\begin{thebibliography}{79}
\bibitem [{sta()}]{state_EP}%
  \BibitemOpen
  \href@noop {} {}\bibinfo {note} {EPs are more abundant than diagonalizable
  degeneracies for generic NH matrices. Specifically, for complex $2\times2$
  matrices with 8 real parameters, EPs form a 6-dimensional manifold compared
  to a 2-dimensional manifold of diagonalizable degeneracies.}\BibitemShut
  {Stop}%
\bibitem [{uns()}]{unsat_case}%
  \BibitemOpen
  \href@noop {} {}\bibinfo {note} {The steady state corresponding to a
  unsatisfiable 3SAT instance is generally a complicated mixed
  state.}\BibitemShut {Stop}%
\end{thebibliography}

\end{document}


\title{Supplemental material for ``Asymptotic Exceptional Steady States in Dissipative Dynamics"}	
	
\author{Yu-Min Hu}
\affiliation{Max Planck Institute for the Physics of Complex Systems, N\"{o}thnitzer Str. 38, 01187 Dresden, Germany}

\author{Jan Carl Budich}
\altaffiliation{jan.budich@tu-dresden.de}
\affiliation{Max Planck Institute for the Physics of Complex Systems, N\"{o}thnitzer Str. 38, 01187 Dresden, Germany}
\affiliation{Institute of Theoretical Physics, Technische Universit\"at Dresden and W\"urzburg-Dresden Cluster of Excellence ct.qmat, 01062 Dresden, Germany}

\maketitle
\section{Brief introduction to the generalized Lindblad master equation}
In this section, we present a brief introduction to the generalized Lindblad master equation [Eq. (1) of the main text]. We will show how  the generalized form arises from weighted ensemble average among different quantum trajectories and discuss possible physical implementations. 

We start with the standard form of the Lindblad master equation [$W=1$ in Eq. (1) of the main text]:
  \begin{equation}\label{seq:std_qme}
          \frac{\mathrm{d}}{\mathrm{d}t}\rho(t)=-i[H,\rho(t)]+  \sum_{\mu}L_\mu\rho (t)L_\mu^\dagger-\frac{1}{2}\{L_\mu^\dagger L_\mu,\rho(t)\}.
  \end{equation}
  This equation describes the time evolution of the density matrix  $\rho$ under the influence of the Hermitian system Hamiltonian $H$  and the quantum jump operators $L_\mu$ caused by system-environment couplings.

This equation can be viewed as an ensemble average in the quantum trajectory picture \cite{daley2014quantum}. A quantum trajectory refers to the stochastic dynamics of a pure state under the influence of non-unitary dynamics and stochastic quantum jumps. We start from an initial state $\ket{\phi(0)}$ at $t=0$.  At each time step $\delta t$, the state $\ket{\phi(t)}$ may evolve to another state $\ket{\phi_{\text{nH}}(t+\delta t)}=\frac{(1-iH_{\text{nH}}\delta t)\ket{\phi(t)}}{\sqrt{p_{\text{nH}}}}$ with the non-Hermitian Hamiltonian $H_{\text{nH}}=H-(i/2)\sum_\mu L_\mu^\dagger L_\mu$ and the normalization factor $p_{\text{nH}}=\braket{\psi(t)|(1-iH_{\text{nH}}\delta t)^\dagger (1-iH_{\text{nH}}\delta t)|\phi(t)}=1-\delta t\sum_\mu\braket{\psi(t)|L_\mu^\dagger L_\mu|\phi(t)}$ . On the other hand, there is also a probability $p_\mu=\delta t \braket{\psi(t)|L_\mu^\dagger L_\mu|\phi(t)}$ such that $\ket{\phi(t)}$ undergoes a quantum jump in the $\mu$-th dissipative channel and becomes a new state $\ket{\phi_\mu(t+\delta t)}=\frac{L_\mu \ket{\phi(t)}}{\sqrt{\braket{\psi(t)|L_\mu^\dagger L_\mu|\phi(t)}}}$. At each time step, $p_{\text{nH}}+\sum_\mu p_\mu=1$ forms a probability distribution conditioned on the state $\ket{\phi(t)}$. Based on this probability distribution, the final state $\ket{\phi(t+\delta t)}$ after this time step is stochastically obtained from these $\ket{\phi_{\text{nH}}(t+\delta t)}$ and $\ket{\phi_\mu(t+\delta t)}$ states. Such a process forms a quantum trajectory undergoing a stochastic dynamics caused by the non-Hermitian Hamiltonian  $H_{\text{nH}}$ and quantum jumps $L_\mu$. The ensemble average of all possible quantum trajectories will lead to the time evolution of the density matrix $\rho(t)$ in Eq. \eqref{seq:std_qme}. 

The ensemble of quantum trajectories can be analyzed using statistical methods. We denote by $\ket{\phi^{(J)}(t)}$ the quantum trajectory that starts at $t=0$ and undergoes $J$ quantum jumps within the time interval $[0, t]$. Here, $J$ represents the total number of quantum jumps that have occurred along this trajectory, which can be measured through tracking the environment dynamics. After a short time step $\delta t$, the system either evolves into a new state $\ket{\phi^{(J)}(t+\delta t)}$ under the non-Hermitian Hamiltonian, or transitions to $\ket{\phi^{(J+1)}(t+\delta t)}$ if a quantum jump occurs. Therefore, we can classify the ensemble of quantum trajectories into several subensembles by the number of quantum jumps in each trajectory. 

Following this classification, we can express the density matrix $\rho(t)$ of the whole ensemble as $\rho(t)=\sum_{J=0}^\infty \rho^{(J)}(t)$ where $\rho^{(J)}(t)$ represents the average over the subensemble labeled by $J$.  We note that this naturally provides a probability distribution of jump numbers $P_J(t)=\operatorname{Tr}[\rho^{(J)}(t)]$ with $\sum_{J=0}^\infty P_J(t)=1=\operatorname{Tr}[\rho(t)]$. The subensemble average $\rho^{(J)}(t)$ follows the time evolution:
\begin{equation}\label{seq:jump_qme}
      \frac{\mathrm{d}}{\mathrm{d}t}\rho^{(J)}(t)=-iH_{\text{nH}}\rho^{(J)}(t)+i\rho^{(J)}(t)H_{\text{nH}}^\dagger+ \sum_{\mu}L_\mu\rho^{(J-1)} (t)L_\mu^\dagger.
\end{equation}
In the above equation, the first term characterizes the  nonunitary dynamics generated by $H_{\text{nH}}$ and the second term (vanishing at $J=0$) represents the occurrence rate of a quantum jump. 

With the distribution $P_J(t)$ of quantum trajectories, it is possible to analyze the counting statistics of $J$ \cite{Garrahan2010Thermodynamics}. The key objective is the generating function $Z_t(s)=\sum_{J=0}^\infty P_J(t)e^{-sJ}$.  This can be viewed as a partition function of a generalized density matrix $\rho_s(t)=\sum_{J=0}^\infty \rho^{(J)}(t)e^{-sJ}$. Combined with Eq. \eqref{seq:jump_qme}, it is easy to obtain the dynamics of $\rho_s(t)$:
\begin{equation}
  \frac{\mathrm{d}}{\mathrm{d}t}\rho_s(t)=-iH_{\text{nH}}\rho_s(t)+i\rho_s(t)H_{\text{nH}}^\dagger+ e^{-s}\sum_{\mu}L_\mu\rho_s (t)L_\mu^\dagger.
  \end{equation}
Defining  $W=e^{-s}$, we obtain the generalized Liouvillian superoperator $\mathcal{L}_W$ defined in Eq. (1) of the main text. It becomes clear that the physical meaning of the parameter $W$ is the weighted average of quantum trajectories.   In this sense, the dynamical instability shown in Fig. 2(a) of the main text can be understood as a biased ensemble average with the enhanced weight for quantum trajectories with more jumps. 

  \section{Classical dynamics in dissipative 3SAT solver}\label{sec:classical}
In this section, we show that the dissipative 3SAT solver discussed in the main text can be reduced to a classical dynamics. In the main text, we consider the Lindblad master equation with $H=0$ and $L_{m,\alpha}=\sigma_{m_\alpha}^xP_m$ where $\alpha=1,2,3$. Given that  $\sigma_{m_a}^x(1\pm \sigma_{m_a}^z)/2=\sigma_{m_\alpha}^\mp$, $L_{m,\alpha}\rho L_{m,\alpha}^\dagger$ only has nontrivial actions on the diagonal part of the reduced density matrix for the qubits involved in clause $C_m$. This indicates that the dissipative dynamics decouples the diagonal and off-diagonal parts of the density matrix. Therefore, in the computational basis where the solution state to a satisfiable instance is just a basis vector, we focus on the classical dynamics occurring between the diagonal elements of the density matrix and focus on the relaxation to the solution state. 

To proceed, we denote the  diagonal operator subspace of the $n$th qubit as
\begin{equation}
    \mid\uparrow_n)=\ket{\uparrow_n}\bra{\uparrow_n},\quad \mid\downarrow_n)=\ket{\downarrow_n}\bra{\downarrow_n}.
\end{equation}
We also denote the classical actions in the diagonal operator subspace as
\begin{equation}
    \begin{aligned}
&\Sigma^x_n\mid\uparrow_n)=\mid\downarrow_n),\quad \Sigma^x_n\mid\downarrow_n)=\mid\uparrow_n),\\
&\Sigma^z_n\mid\uparrow_n)=\mid\uparrow_n),\quad \Sigma^z_n\mid\downarrow_n)=-\mid\downarrow_n).\\
    \end{aligned}
\end{equation}
With these notations, we represent the diagonal elements of the density matrix $\rho$ in the qubit computation basis as a classical probability distribution among the corresponding classical bit strings:
\begin{equation}
    p_i=\braket{i|\rho|i}\in[0,1],
\end{equation}
where $\ket{i}$ takes from $2^N$ computational basis vectors. They form a $2^N$-component real vector $\vec{p}$.  As  a result, the Lindblad master equation for Eq.(3) of the main text leads to the vector $\vec{p}$ following a classical evolution $\frac{\mathrm{d}\vec{p}}{\mathrm{d}t}=\mathcal{M}_W\vec{p}$ with the generator given by
\begin{equation}\label{eq:classical_M}
    \mathcal{M}_W=\sum_{m=1}^M\sum_{\alpha=1}^3(W\Sigma_{m_\alpha}^x-1)\mathcal{P}_m.
\end{equation}
Here, $\mathcal{P}_m$ is the analogous projector on the diagonal basis. For example, $C_m=x_{m_1}\lor \neg {x}_{m_2}\lor x_{m_3}$ corresponds to $\mathcal{P}_m=(1-\Sigma_{m_1}^z)(1+\Sigma_{m_2}^z)(1-\Sigma_{m_3}^z)/8$. The physical classical Markovian dynamics is given by $W=1$, where the sum of each column of $\mathcal{M}_1$ is equal to zero and the total probability $\sum_{i=1}^{2^N} p_i=\text{Tr}[\rho]=1$ is conserved. 

The classical dynamical generator $\mathcal{M}_W$ allows us to obtain numerical results for a relatively large system size. Practically, we exactly diagonalize $\mathcal{M}_W$ to obtain the numerical data in Figs. 1 and 2 of the main text. 
 
\section{Generating satisfiable 3SAT instances }
The numerical results in the main text require the generation of random satisfiable and hard-to-solve  3SAT instances. Given the number of variables $N$, we generate $M=\text{round}(\alpha_cN)$ clauses for each instance. The parameter $\alpha_c\approx4.267$ is the satisfiability threshold for the 3SAT problem \cite{Mezard2002random}. Below this critical value, most random 3SAT instances tend to be satisfiable; above it, they are overwhelmingly unsatisfiable. The instances near this critical value are the most computationally challenging ones.

In this paper, we require all the generated 3SAT instances to be satisfiable. This is done by the method developed in Ref. \cite{Barthel2002Hiding} with a parameter $p_0=0.08$. This method generates satisfiable hard instances with at least one solution. We then only select instances with a unique solution for the numerical investigation in the main text. We also keep instances with two solutions for the numerical investigation in Sec. \ref{sec:twosol} of this supplemental material. 

Before closing this section, we remark that the specific instances used in Fig. 1, Fig. 2(a,b,c), and Fig.  3(b) in the main text are employed as examples. The critical values $W_c$ in Fig. 2(c) and Fig. 3(b) are also determined for each instance. Nevertheless, the instance-dependent results shown in these plots are qualitatively the same for different instances. 

\section{Eigenoperator structures of AESS}

\begin{figure*}[t]
    \centering
    \includegraphics[width=\textwidth]{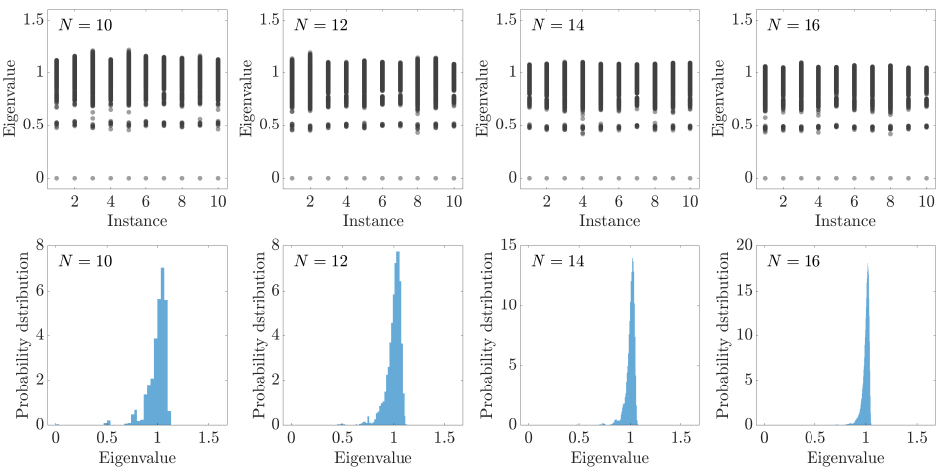}
    \caption{The spectrum of left eigenoperator $l_1$ for the classical dissipative 3SAT solver. Top: The eigenvalues of $\kappa^{-1}l_1$ for ten typical satisfiable 3SAT instances with a unique solution. Bottom: the probability distribution function of eigenvalues of the first instance in the top row.   Each column corresponds to a specific variable number $N$, which is shown in each plot. We consider the physical case $W=1$.}
    \label{sfig:left}
\end{figure*}

\begin{figure}[t]
    \centering
    \includegraphics[width=8cm]{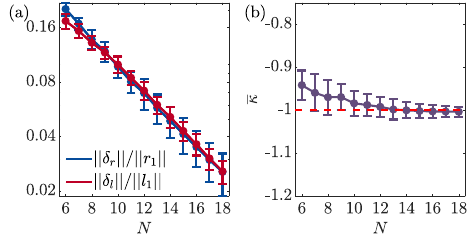}
    \caption{The small corrections of  $l_1$ and $r_1$  in the classical dissipative 3SAT solver. (a) The averaged relative norm of the small corrections $\delta$ in $r_1$ and $l_1$. (b) The mean value of $\kappa$. The red dashed line represents $\kappa=-1$. Each point is averaged over $10^3$satisfiable instances, with the error bar implying the standard deviation among different instances. Each instance is selected to have one unique solution. We consider the physical case $W=1$.}
    \label{sfig:correction}
\end{figure}

In this section, we analyze the generic eigenoperator structure of $l_1$ and $r_1$ of AESS. In dissipative state preparation, the target steady state $r_0=\ket{\psi_{\text{sol}}}\bra{\psi_{\text{sol}}}$ is a projector and $l_0=I$ is the identity. The biorthonormal relations $\text{Tr}[l_0^\dagger r_1]=\text{Tr}[l_1^\dagger r_0]=0$ give rise to $\text{Tr}[r_1]=0$ and $\braket{l_1}=0$, where we define $\braket{\cdot}=\text{Tr}[r_0^\dagger (\cdot)]=\braket{\psi_{\text{sol}}|\cdot|\psi_{\text{sol}}}$.  The presence of AESS indicates that $r_1$ is close to  $r_0$. We thus decompose $r_1$ as $r_1=\braket{r_1}r_0+\delta_r$. The deviation $\delta_r$ is orthogonal to $r_0$ under Hilbert-Schmidt inner product, satisfying $\braket{\delta_r}=0$. The biorthonormal relation $\text{Tr}[l_1^\dagger r_1]=1$ leaves freedom to choose the normalization $||r_1||$. We take $\braket{r_1}=1$, which leads to $r_1=r_0+\delta_r$ with $\text{Tr}[\delta_r]=-1$. These properties lead to $||\delta_r||^2=\text{Tr}[\delta_r^\dagger\delta_r]=1-|\nu|^{-2}$ with $\nu$ defined in the main text, indicating that the correction $\delta_r$ (and its norm) asymptotically vanishes in a large system with AESS.  Similarly for $l_1$, provided that $\braket{l_1^\dagger}=0$, we take the ansatz $l_1=\kappa(I-r_0) +\delta_l$ where $\braket{\delta_l}=\text{Tr}[\delta_l]=0$ and $\kappa=\text{Tr}[l_1]/(2^N-1)$. The relation $\text{Tr}[l_1^\dagger r_1]=1$ leads to $\text{Tr}[{\delta_l^\dagger \delta_r}]=1+\kappa$.  We then calculate the eigenoperator overlap between $l_0$ and $l_1$: $|v^\prime|^2=\frac{|\text{Tr}[l_0^\dagger l_1]|^2}{||l_0||^2|| l_1||^2}=(1-\frac{1}{2^N})(1-\frac{||\delta_l||^2}{||l_1||^2})$. Since an AESS indicates $|v^\prime|^2\to1$, we get $\frac{||\delta_l||^2}{||l_1||^2}\to0$ as the system size increases. Thus, $||\delta_l||$ may also be viewed as a small correction relative to the extensive operator $\kappa(I-r_0)$.

Our extensive numerical investigations of the dissipative 3SAT solver further support the above discussion. We first show the numerical results of the eigenvalues of $\kappa^{-1}l_1$ for different instances [Fig. \ref{sfig:left}]. Given that the eigenvalue $\lambda_1$ of the metastable state is observed to be real, the left eigenoperator $l_1$ is Hermitian \cite{Minganti2018spectral} and thus has a real spectrum. We find that there is a single zero eigenvalue of each $\kappa^{-1}l_1$, whose eigenstate corresponds to $\ket{\psi_{\text{sol}}}$. Other eigenvalues of $\kappa^{-1}l_1$ are positive and have a constant gap above a single zero eigenvalue.  These results indicate that $\kappa^{-1}l_1$ is positive semidefinite. Interestingly, except for a few modes,  most eigenvalues are centered around $1$, indicating that $\kappa^{-1}l_1$ is close to the extensive projector $I-r_0$.

Furthermore, Fig. \ref{sfig:correction}(a) shows that the mean value of $\frac{||\delta_l||}{||l_1||}$ and $\frac{||\delta_r||}{||r_1||}$ among different satisfiable 3SAT instances exhibits an exponential decay, which is consistent with the above analysis. Additionally, Fig. \ref{sfig:correction}(b) demonstrates that $\kappa=\text{Tr}[l_1]/(2^N-1)$ converges to $-1$ with the increase of variable number $N$. With all the numerical results in Figs. \ref{sfig:left} and  \ref{sfig:correction}, we conclude that $l_1$ is very close to $r_0-I$ in the operator space.  In this sense, for a random initial state $\rho_{\text{ini}}$ without prior knowledge of the target state, we have $c_1 = \text{Tr}[l_1^\dagger \rho_{\text{ini}}] = -1 + \epsilon$, where $\epsilon$ is typically an exponentially small correction. 

\section{Deriving Eq.(4) of the main text}
We here present the derivation of Eq.(4) in the main text. 

We note that the normalized right eigenstates $\hat r_0$ and $\hat r_1$ satisfy the following eigenequation for the Liouvillian superoperator $\mathcal{L}_W$:
\begin{equation}
    \mathcal{L}_W[\hat r_0]=0,\quad \mathcal{L}_W[\hat r_1]=\lambda_1\hat r_1.
\end{equation}
Here, $\hat r_0$ and $\hat r_1$ are normalized under the Hilbert-Schmidt norm such that $\operatorname{Tr}[\hat r_0^\dagger \hat r_0]=\operatorname{Tr}[\hat r_1^\dagger \hat r_1]=1$. We note that the first equation always holds for arbitrary $W$ if the target state $\hat r_0$ is encoded in the common decoherence-free subspace of all jump operators, like the examples considered in the main text (see Fig.2(a) of the main text). We also define $\nu=\operatorname{Tr}[\hat r_0^\dagger\hat r_1]$, which is the Hilbert-Schmidt inner product between these two normalized operators. If $\hat r_0$ and $\hat r_1$ are linearly independent, we can perform the Gram-Schmidt orthogonalization of these two operators to obtain an orthonormal operator basis spanned by $\tilde r_0$ and $\tilde r_1$:
\begin{equation}
    \tilde r_0=\hat r_0, \quad \tilde r_1=\frac{\hat r_1-\nu \hat r_0}{(1-|\nu|^2)^{1/2}}.
\end{equation}
It is straightforward to check that  $\operatorname{Tr}[\tilde r_0^\dagger \tilde r_0]=\operatorname{Tr}[\tilde r_1^\dagger \tilde r_1]=1$ and $\operatorname{Tr}[\tilde r_0^\dagger \tilde r_1]=0$. Within this operator subspace, the action of the Liouvillian superoperator $\mathcal{L}_W$ becomes
\begin{equation}
\begin{split}
    \mathcal{L}_W[\tilde r_0]&=0,\\
    \mathcal{L}_W[\tilde r_1]&=\frac{\lambda_1}{(1-|\nu|^2)^{1/2}}\hat r_1=\lambda_1\tilde r_1+\frac{\nu  \lambda_1}{(1-|\nu|^2)^{1/2}}\tilde r_0.
    \end{split}
\end{equation}
As a result, within the two-dimensional operator subspace spanned by the two orthonormal basis operators  $\tilde r_0$ and $\tilde r_1$, the effective Liouvillian superoperator becomes a 2-by-2 matrix:
\begin{equation}
\mathcal{L}_{W}^{\text{eff}}=\begin{pmatrix}
        0 & \lambda_1 \frac{\nu}{\sqrt{1-|\nu|^2}} \\ 0 & \lambda_1
    \end{pmatrix}.
\end{equation}
This is exactly Eq. (4) of the main text. 

\section{Degenerate steady states}\label{sec:twosol}

In the main text, we focus on the asymptotic exceptional steady state (AESS) in classical and quantum systems that have a unique steady state. This section demonstrates that AESS can coexist with multiple steady states, and that our focus on uniquely solvable 3SAT instances in the main text does not lead to a loss of generality. 

When a Liouvillian $\mathcal{L}_{W=1}$  has more than one steady state, the spectral decomposition $\mathcal{L}_1[r_i]=\lambda_i r_i$ and $\mathcal{L}_1^\dagger[l_i]=\lambda_i^* l_i$  gives rise to $\lambda_0=\lambda_1=\cdots=\lambda_{d-1}=0$ where $d$ is the degeneracy of steady states. The $d$ eigenoperators $\{r_0,r_1,\cdots,r_{d-1}\}$ in the steady-state subspace of $\mathcal{L}_1$ are linearly independent and form a $d$-dimensional subspace \cite{Minganti2018spectral}. Namely, they cannot form any EP subspace by themselves. We can employ the Gram-Schmidt decomposition to construct an orthonormal operator basis based on these $d$ eigenoperators. For simplicity, we assume that this step has been done and $\{r_0,r_1,\cdots,r_{d-1}\}$ are orthogonal to each other. 

Other modes correspond to $0>\text{Re}(\lambda_d)\ge\text{Re}(\lambda_{d+1})\ge\text{Re}(\lambda_{d+2})\ge\cdots$. We do not consider the situation with purely imaginary eigenvalues, although the analysis below can also apply to that case by tracking the eigenmodes whose eigenvalue asymptotically approaches zero in the thermodynamic limit. 

Similar to the main text, we define the Liouvillian gap as 
\begin{equation}
    \Delta=-\text{Re}(\lambda_d).
\end{equation} 
The eigenstate overlap is defined between the metastable state $r_d$ and the steady-state subspace spanned by the orthogonal basis $\{r_0,r_1,\cdots, r_{d-1}\}$. With the normalized  eigenoperator $\hat r_i=r_i/||r_i||$, we have the following definition:
\begin{equation}
    |\nu|^2=\sum_{k=0}^{d-1}|\text{Tr}[\hat r_d^\dagger\hat r_k]|^2.
\end{equation}

Therefore, we can expect that the AESS coexists with multiple steady states when both $\Delta$ and $1-|\nu|^2$ approach zero as the system size increases.  In this case, the metastable state can have a large overlap with a particular state in the steady-state subspace, while keeping orthogonal to other states.  

\begin{figure}[t]
    \centering
    \includegraphics[width=8.5cm]{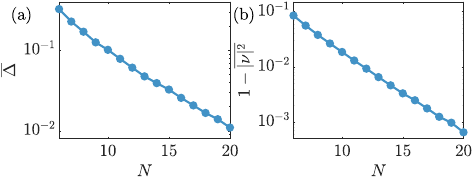}
    \caption{The mean dissipative gap $\overline{\Delta}$ and eigenoperator overlap $1-\overline{|\nu|^2}$  at $W=1$ for 3SAT instances with two distinct satisfiable solutions. Each data point is obtained by diagonalizing $\mathcal{M}_1$ and averaging over $10^3$ random instances. For each variable number $N$,  $M=\text{round}(\alpha_cN)$ and $\alpha_c\approx4.267$.}
    \label{sfig:twosol}
\end{figure}

As an example, we consider the classical dynamics introduced in Eq.\eqref{eq:classical_M} for 3SAT instances with two different satisfiable solutions $\ket{\psi_{\text{sol},1}}$ and $\ket{\psi_{\text{sol},2}}$. By definition, $\ket{\psi_{\text{sol},1}}$ and $\ket{\psi_{\text{sol},2}}$ are just two classical states in the computational basis to represent the two bit configurations that solve the given instance.  Since our classical dynamics only involves the diagonal parts of the density matrix, the corresponding two steady states of $\mathcal{M}_1$ are $r_0=\ket{\psi_{\text{sol},1}}\bra{\psi_{\text{sol},1}}$ and $r_1=\ket{\psi_{\text{sol},2}}\bra{\psi_{\text{sol},2}}$ \footnote{A full quantum consideration indicates that $\mathcal{L}_1$ has four steady states.  The other two are off-diagonal: $r_2=\ket{\psi_{\text{sol},1}}\bra{\psi_{\text{sol},2}}$ and $r_3=\ket{\psi_{\text{sol},2}}\bra{\psi_{\text{sol},1}}$. } Therefore, $\Delta=-\text{Re}(\lambda_2)$ and $|\nu|^2=|\braket{\psi_{\text{sol},1}|\hat r_2|\psi_{\text{sol},1}}|^2+|\braket{\psi_{\text{sol},2}|\hat r_2|\psi_{\text{sol},2}}|^2$. The mean quantities $\overline{\Delta}$ and $1-\overline{|\nu|^2}$, averaged over $10^3$ random 3SAT instances with two satisfiable solutions, are shown in Fig.\ref{sfig:twosol}. These two quantities display an exponential decay as the system size increases, similar to Fig.1 of the main text. These results demonstrate that the AESS can coexist with multiple steady states. 

\begin{figure}[t]
    \centering
    \includegraphics[width=8.5cm]{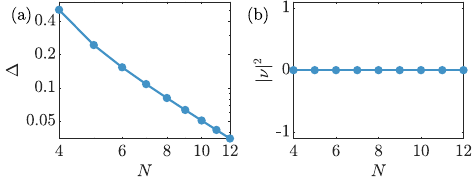}
    \caption{The Liouvillian gap $\Delta$  (a) and the eigenoperator overlap $|\nu|^2$ (b) for the dephasing XX chain $\mathcal{L}_1$ defined in Sec. \ref{sec:eg1}. }
    \label{sfig:xx}
\end{figure}

\section{Examples without asymptotic exceptional steady states}

In the main text and the previous section of the supplemental material, we mainly focus on classical and quantum systems with AESS. In this section, we show two examples without AESS, where only one of the limits $\lim_{N\to\infty}\Delta=0$ and $\lim_{N\to\infty}|\nu|^2=1$ can be satisfied in the thermodynamic limit. 
\subsection{$\lim_{N\to\infty}\Delta=0$ and $\lim_{N\to\infty}|\nu|^2\ne1$}\label{sec:eg1}

We use a dephasing XX chain as an example in this case \cite{Marko2015Relaxation}. The Hamiltonian is given by $ H=\sum_{n=1}^{N-1}s_n^xs_{n+1}^x+s_{n}^ys_{n+1}^y$ and the jump operators are dephasing operators  $L_n=s_n^z$. We take a unit damping rate. Here we consider open boundary conditions for a spin-$\frac{1}{2}$ chain with $N$ spins. The total Liouvillian at $W=1$ is given by $\mathcal{L}_1[\rho]=-i[H,\rho]+\sum_{n=1}^N L_n\rho L_n^\dagger-\frac{1}{2}\{L_n^\dagger L_n,\rho\}$.  This open quantum system has a strong  global $U(1)$ symmetry such that the total magnetization $S^z=\sum_{n=1}^Ns_n^z$ is conversed.  Meanwhile, since $L_n=L_n^\dagger$ is a Hermitian operator, the steady state for $\mathcal{L}_1$ is the identity operator in each symmetry sector labeled by the eigenvalues of $S^z$. 

The eigenvalues and eigenoperators of $\mathcal{L}_1$ can be obtained in each symmetry sector. We perform exact diagonalization in the sector $S^z=0$ for even spins and in the sector $S^z=-\frac{1}{2}$ for odd spins. Fig. \ref{sfig:xx} shows the numerical data for the Liouvillian gap $\Delta$ and eigenoperator overlap $\nu$ between the steady state and the slowest damping mode in these sectors. These results show that the Liouvillian gap algebraically decays to zero with the increase of the system size. However, the eigenoperator overlap $\nu$ remains zero, indicating that the steady state and the slowest damping mode are orthogonal to each other under Hilbert-Schmidt inner product. The zero overlap is expected here. The right eigenoperators of $\mathcal{L}_1$ with nonzero eigenvalues are traceless, and therefore, orthogonal to the steady state which is an identity matrix. The results in Fig. \ref{sfig:xx} suggest that a gapless Liouvillian with Hermitian jump operators cannot yield an AESS in the thermodynamic limit.

\begin{figure}[t]
    \centering
    \includegraphics[width=8.5cm]{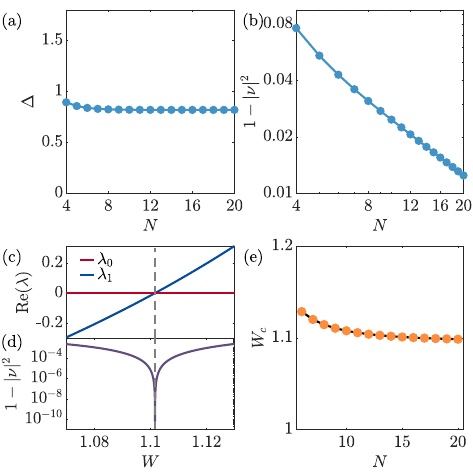}
    \caption{The Liouvillian gap $\Delta$  (a) and the eigenoperator overlap $|\nu|^2$ (b) for a classical spin chain in Sec. \ref{sec:eg2} at the physical point $W=1$. (c) and (d) show the $W$-dependent Liouvillian gap $\Delta$ and eigenoperator overlap $|\nu|^2$ for $N=14$. an exact EP appears at the critical value $W_c=1.10167$. (e) The critical value $W_c$ for different system sizes $N$. The black curve is fitted by $W_c=1.0546N^{-1.9281}+1.0952$, indicating a finite distance to the physical point $W=1$ in the thermodynamic limit.}
    \label{sfig:Ising}
\end{figure}

\subsection{$\lim_{N\to\infty}\Delta\ne 0$ and $\lim_{N\to\infty}|\nu|^2=1$}\label{sec:eg2}

The second example is a 1D classical Markovian dynamics for $N$ classical bits. The generator is $\mathcal{M}_W=\sum_{n=1}^{N}(W\Sigma^x_n+W\Sigma^x_{n+1}-2)(1-\mathcal{P}_{n,n+1}^{\downarrow\downarrow})$. These notations are taken from Sec.\ref{sec:classical} of this supplemental material. $\mathcal{P}_{n,n+1}^{\downarrow\downarrow}=\mid\downarrow\downarrow)(\downarrow\downarrow\mid $ is a projection superoperator of two aligned neighboring spins in the downward direction. We consider the periodic boundary condition here. By the construction, the steady state of this classical Markovian dynamics is given by the classical ferromagnetic state where all spins are downwards. It is a trivial task to prepare such a state. 

Fig. \ref{sfig:Ising} shows the dissipative gap $\Delta$ and the eigenoperator overlap $\nu$ between the steady state and the slowest decay mode of $\mathcal{M}_1$. The dissipative gap $\Delta$ is nearly constant, not decaying as the system size increases. In contrast, $1-|\nu|^2$ exhibits a power-law decay with the increase of the system size. According to $\mathcal{L}_W^{\text{eff}}$ in the main text, we can write down an effective dynamical generator in the orthonormal basis constructed by the steady state and the slowest decay mode of $\mathcal{M}_1$: 
\begin{equation}
    \mathcal{M}_1^{\text{eff}}=\begin{pmatrix}
        0 & -\Delta \frac{\nu}{\sqrt{1-|\nu|^2}} \\ 0 &-\Delta
    \end{pmatrix}.
\end{equation}
A finite gap $\Delta\ne0$ indicates that $\mathcal{M}_1^{\text{eff}}$ will not asymptotically become an EP as the system size increases. Nevertheless, the two eigenoperators asymptotically become parallel to each other in the thermodynamic limit. We can express $\mathcal{M}_1^{\text{eff}}$ as 
\begin{equation}\label{seq:effM}
    \mathcal{M}_1^{\text{eff}}=-\Delta \frac{\nu}{\sqrt{1-|\nu|^2}}\begin{pmatrix}
        0 & 1 \\ 0 & \frac{\sqrt{1-|\nu|^2}}{\nu}
    \end{pmatrix}.
\end{equation}
As the system size increases, this matrix looks like an asymptotic Jordan block multiplied with a divergent prefactor. 

We stress that this situation differs from AESS in the main  text. The latter case, manifesting as both $\Delta$ and $1-|\nu|^2$ vanishing in the thermodynamic limit, stays close to an exact EP at a critical value $W_c>1$.  The critical value $W_c$ for AESS also converges to the physical point $W=1$ in the thermodynamic limit. In contrast, although $\mathcal{M}_1^{\text{eff}}$ in Eq. \eqref{seq:effM} is also found to be near an exact EP at $W=W_c>1$ [Figs. \ref{sfig:Ising}(c) and  \ref{sfig:Ising}(d)], the critical value $W_c$ remains a finite distance to the physical point $W=1$ with the increase of the system size [Fig.  \ref{sfig:Ising}(e)].

\bibliography{AEP_ref}